\documentclass[aps,twocolumn,amssymb,prl,showpacs,nofootinbib,10pt]{revtex4-1}

\usepackage[utf8]{inputenc}
\usepackage{tabularx}
\usepackage{multirow}
\usepackage{url}
\usepackage{amsmath}
\usepackage{graphicx}
\usepackage{mathrsfs}
\usepackage{color}
\usepackage[normalem]{ulem}
\usepackage{float}
\usepackage{lmodern}
\usepackage{soul}
\usepackage[breaklinks,colorlinks,urlcolor=blue,citecolor=blue]{hyperref}

\usepackage{amsmath}
\usepackage{mathrsfs}
\allowdisplaybreaks

 \newcommand{\araa}{ARA\&A}%
\newcommand{\apjl}{ApJ}%
\newcommand{\ssr}{Space~Sci.~Rev.}%
\newcommand{\aapr}{A\&ARv}

\def\simlt{\lower.5ex\hbox{$\; \buildrel < \over \sim \;$}}
\def\simgt{\lower.5ex\hbox{$\; \buildrel > \over \sim \;$}}

\def\beq{\begin{equation}}
\def\eeq{\end{equation}}
\def\ba{\begin{eqnarray}}
\def\ea{\end{eqnarray}}
\def\bB{\boldsymbol{B}}
\def\bE{\boldsymbol{E}}

\def\bv{\boldsymbol{v}}

\def\Eq{Eq.}
\def\Eqs{Eqs.}

\def\sT{\sigma_{\rm T}}

\def\E{{\cal E}}

\def\omL{\omega_{\rm L}}

\def\omB{\omega_B}

\def\Bbg{B_{\rm bg}}

\def\bBbg{\bB_{\rm bg}}

\def\xirise{\xi_{\rm rise}}

\def\bux{\bar{u}_x}

\def\bg{\bar{\gamma}}
\def\baa{\overline{a^2}}
\def\bu{\bar{u}}

\def\vph{v_{\rm ph}}

\def\omsc{\omega_c}

\def\gav{\langle\gamma\rangle}

\def\ssc{\sigma_{\rm sc}}
\def\vu{\boldsymbol{u}}
\def\bb{\boldsymbol{\beta}}
\def\bomB{\boldsymbol{\omega}_B}

\def\epsc{\epsilon_c}

\def\dEem{\dot{\E}_{\rm em}}
\def\dgem{\dot{\gamma}_{\rm em}}

\def\bfrr{\boldsymbol{f}}

\def\gL{\gamma_{\rm RRL}}

\def\tL{t_{\rm L}}
\def\xiL{\xi_{\rm L}}

\def\gs{\gamma_\star}

\def\as{a_\star}

\def\omw{\omega_{\rm w}}

\newbox\grsign \setbox\grsign=\hbox{$>$} \newdimen\grdimen \grdimen=\ht\grsign
\newbox\simlessbox \newbox\simgreatbox \newbox\simpropbox
\setbox\simgreatbox=\hbox{\raise.5ex\hbox{$>$}\llap
     {\lower.5ex\hbox{$\sim$}}}\ht1=\grdimen\dp1=0pt
\setbox\simlessbox=\hbox{\raise.5ex\hbox{$<$}\llap
     {\lower.5ex\hbox{$\sim$}}}\ht2=\grdimen\dp2=0pt
\setbox\simpropbox=\hbox{\raise.5ex\hbox{$\propto$}\llap
     {\lower.5ex\hbox{$\sim$}}}\ht2=\grdimen\dp2=0pt
\def\simgt{\mathrel{\copy\simgreatbox}}
\def\simlt{\mathrel{\copy\simlessbox}}

\begin{document}

\title{Scattering of ultrastrong electromagnetic waves by magnetized particles}

\author{Andrei M. Beloborodov$^{1,2}$}

\affiliation{$^1$Physics Department and Columbia Astrophysics Laboratory, Columbia University, 538 West 120th Street, New York, NY 10027, USA}

\affiliation{$^2$Max Planck Institute for Astrophysics, Karl-Schwarzschild-Str. 1, D-85741, Garching, Germany}

\begin{abstract}
Observations of powerful radio waves from neutron star magnetospheres raise the question of how strong waves interact with particles in a strong background magnetic field $\Bbg$. This problem is examined by solving the particle motion in the wave. Remarkably, waves with amplitudes $E_0>\Bbg$ pump particle energy via repeating resonance events, quickly reaching the radiation reaction limit. As a result, the wave is scattered with a huge cross section. This fact has implications for models of fast radio bursts and magnetars. Particles accelerated in the wave emit gamma-rays, which can trigger an $e^\pm$ avalanche and, instead of silent escape, the wave will produce X-ray fireworks.
\end{abstract}

\maketitle


{\em Introduction.---}
Magnetized compact objects are capable of generating strong electromagnetic waves of low frequencies $\omega$. In particular, pulsars and magnetars produce bright radio emission, and magnetars are thought to be the sources of fast radio bursts (FRBs) \citep{Kaspi17,Petroff19}. 

A strong background magnetic field $\bBbg$ is believed to suppress the plasma response to the wave electric field $\bE$ when $\bE\perp\bBbg$. This allows the linearly polarized radio waves to freely propagate through the magnetospheric plasma even when $\omega$ is far below the plasma frequency. A standard calculation of the cross section for wave scattering by a magnetized electron gives $\ssc\approx (\omega/\omega_B)^2\sT\ll \sT$ \citep{Canuto71}, where $\omega_B$ is the gyrofrequency and $\sT$ is the Thomson cross section.

However, the standard analysis fails for waves with amplitudes $E_0>\Bbg$, and this regime is inevitably encountered as a strong wave packet  propagates away from the magnetized star, in the decreasing $\Bbg$. Note that both conditions $\Bbg<E_0$ and $\omega_B>\omega$ may be expressed as
\beq
\label{eq:cond}
   1<\frac{\omega_B}{\omega}<a_0, \qquad a_0=\frac{eE_0}{mc\omega},
\eeq
where $e$ and $m$ are the electron charge and mass. Low-frequency waves can have enormous $a_0$. For instance, FRB models picturing a GHz source of luminosity $L\sim 10^{42}\,$erg/s at radii $R<10^8$\,cm have $E_0^2=2L/cR^2$ and $a_0\sim 10^5 R_8^{-1}$. The wave encounters regime~(\ref{eq:cond}) in the outer magnetosphere $R\gtrsim  3\times 10^8\,$cm where $\Bbg\propto R^{-3}$ drops below $E_0$ \citep{Beloborodov21}. We find below that regime~(\ref{eq:cond}) triggers quick stochastic acceleration of particles in the wave, and $\ssc$ becomes huge.

\medskip

{\em Method.---}
Let us consider a wave packet  propagating along  $\hat{\boldsymbol{z}}$ in an initially static magnetized plasma of low density, and calculate the particle motion in the packet. It obeys the dynamical equation for velocity $\bv=\bb c$ or four-velocity $u^\mu=(\gamma,\vu)$ (where $\vu=\gamma\bb$):
\beq
\label{eq:dyn}
    mc\,\frac{d\vu}{dt}=e\left[\bE + \bb\times(\bB+\bBbg)\right] + \bfrr,
\eeq
where $\bfrr$ is the radiation reaction force. Relevant scales in this dynamical problem are microscopic compared with the scale $R$ of the background field variation, so $\bBbg$ can be approximated as uniform.
The simple case of $\bBbg=0$ has been extensively studied in laser plasma physics  \citep{Bulanov15}. The particle motion was also solved for circularly polarized waves propagating along $\bBbg\neq 0$ \citep{Zeldovich7Z}. In these cases, the stochastic pump effect described below disappears. 

The wave fields $\bE$ and $\bB$ depend on $\xi=t-z/c$ and described by the dimensionless potential $\boldsymbol{a}=e\boldsymbol{A}/mc^2=(a(\xi),0,0)$:
\beq
\label{eq:EB}
  \frac{e\bE}{mc}=\left(-\frac{da}{d\xi},0,0\right), \qquad 
  \frac{e\bB}{mc}=\left(0,-\frac{da}{d\xi},0\right).
\eeq
We choose $\xi=0$ at the leading edge of the wave packet, so it propagates at $\xi>0$ (and $a=0$ at $\xi\leq 0$). In numerical examples we will use a modulated sine wave with amplitude rising linearly at $0<\xi<\xirise$ and then staying constant in the packet: $E(\xi)=E_0\sin(\omega\xi)$. 

Consider a particle initially at rest before the wave ($\bb=0$ at $\xi\leq 0$). The wave overtakes the particle with relative speed $d\xi/dt=1-\beta_z$, and we define
\beq
\label{eq:dtau}
   u_\xi\equiv \frac{d\xi}{d\tau}  =\gamma(1- \beta_z)=\gamma-u_z, 
   \qquad d\tau\equiv\frac{dt}{\gamma}.
\eeq
Note that the wave potential $a$ varies along the particle worldline with  $da/d\tau=-(eE/mc)u_\xi$. We also define
\beq
  \bomB=\frac{e\bBbg}{mc}=(0,\omB\sin\theta,\omB\cos\theta), 
  \qquad \theta\neq 0.
\eeq
The case of $\theta=\pi/2$ is particularly simple --- then the wave does not excite $u_y$, i.e. $\vu=(u_x,0,u_z)$.

We first examine particle motion without radiation reaction, $\bfrr\approx 0$. 
Then, \Eq~(\ref{eq:dyn}) gives
\beq
\label{eq:dyn1}
    \frac{dU_x}{d\tau} =  \omB^z u_y - \omB^y u_z, \;
    \frac{du_\xi}{d\tau} =-\omB^y u_x, \;
    \frac{du_y}{d\tau} = \omB^z u_x,
\eeq
where $U_x=u_x+a$. Variables ($U_x$,$u_\xi$,$u_y$) determine all components of $u^\mu$ (using $u^\mu u_\mu=-1$). We solve \Eqs~(\ref{eq:dyn1}) with initial conditions $\vu=0$, $U_x=0$, $u_\xi=1$ at $\xi=0$.

When $\Bbg=0$, the solution is trivial: $U_x$ and $u_\xi$ keep their initial values, which yields
\beq
\label{eq:umu_Bbg_zero}
   u_x=-a, \quad  u_z=\frac{a^2}{2}, \quad \gamma=1+\frac{a^2}{2}.
\eeq
This motion is well known, although it is usually viewed in the center-of-momentum frame $K'$ where the average $\bu_z'=0$ and the particle executes an 8-shaped orbit \citep{Landau75}. 

For waves with $E_0\ll \Bbg$, the particle motion is also known: it oscillates in the wave with small $|\vu|\sim E_0/\Bbg$. 

Hereafter we focus on waves with $E_0>\Bbg\neq 0$. Then, we find that the particle motion is a superposition of fast $\omega$-oscillations in the wave with amplitude $|\Delta\vu|\sim a_0$ and a slower Larmor rotation of $\bar\vu$ (averaged over the $\omega$-oscillations) in the average field $\overline{\bB+\bBbg}=\bBbg$. In particular, $\Delta u_x=a_0$ and $\bu_x=U_x$.

\begin{figure*}[t]
\includegraphics[width=0.63\textwidth]{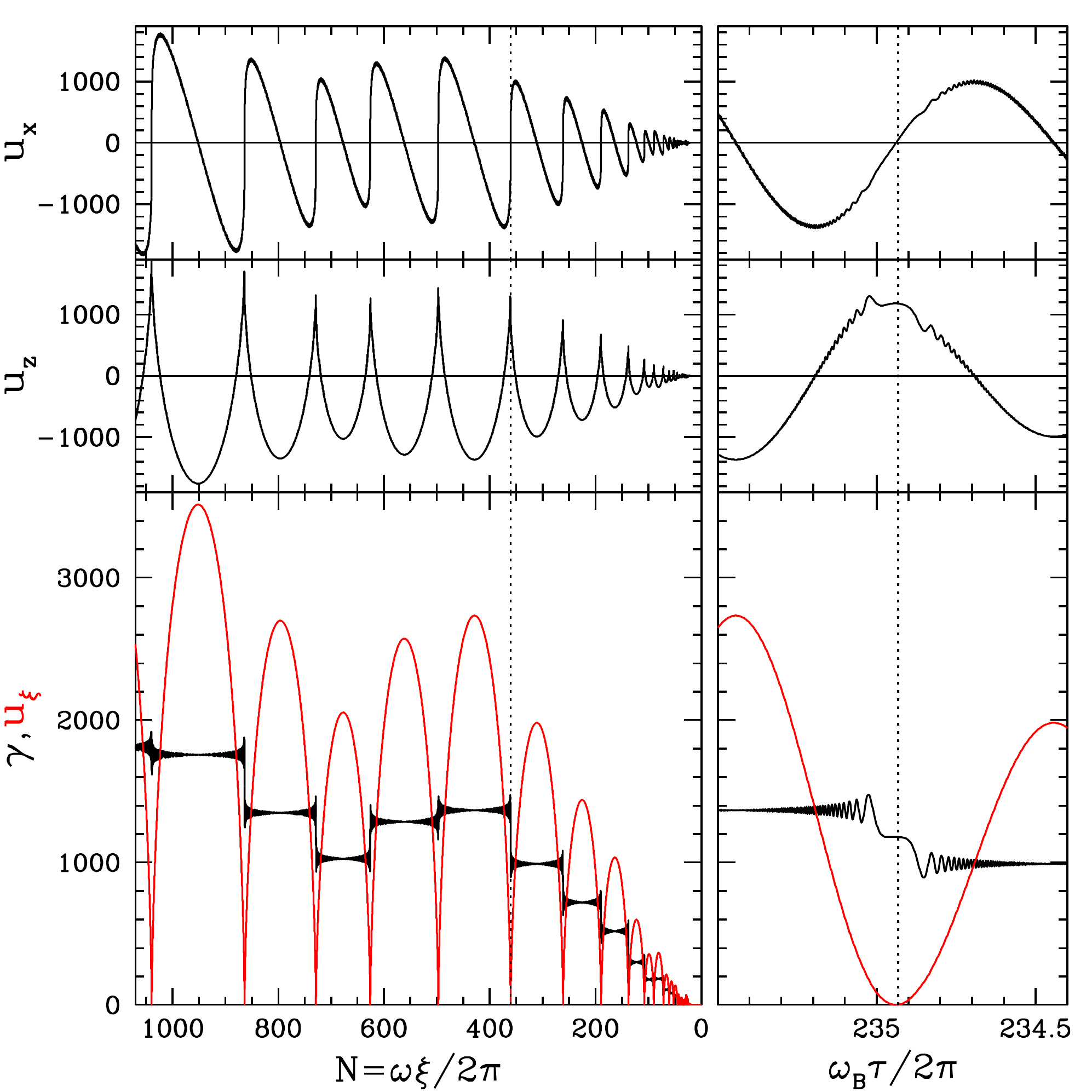} 
\hspace*{-5cm}
\includegraphics[width=0.63\textwidth]{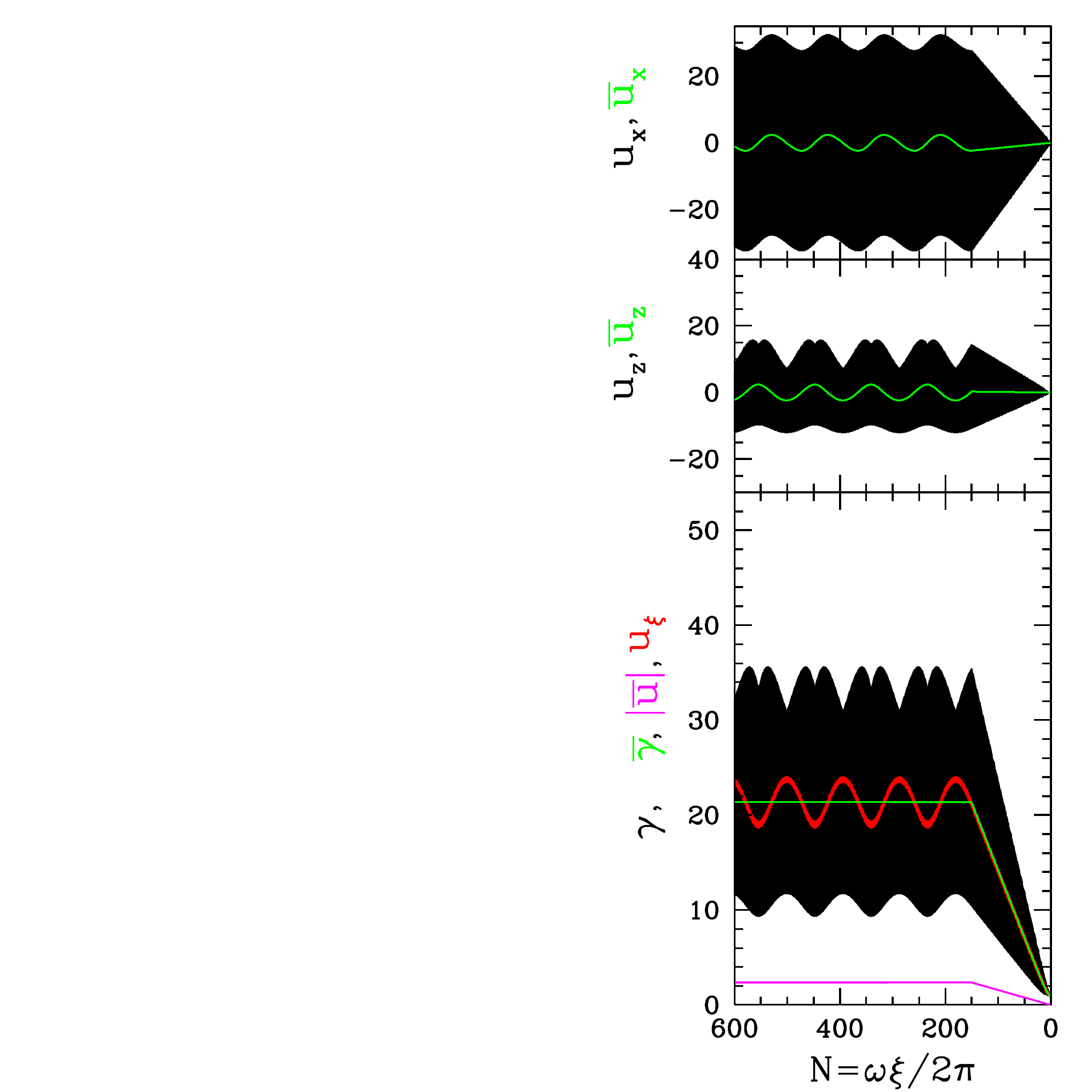} 
\caption{
Two types of particle motion in strong waves ($E_0>\Bbg$), illustrated using a wave with $a_0=30$ and $\xirise=150(2\pi/\omega)$. The leading edge of the wave packet is at $\xi=0$; the particle's coordinate $\xi=t-z/c$ grows with time to the left.
Left: $\omega_B=10\omega$. One can see large $|\bar\vu|\gg|\Delta\vu|$ pumped by resonances that happen every Larmor time $\xiL\approx \tL=2\pi/\omL$. One of the resonances is indicated by the vertical dotted line. The behavior of $u^\mu$ near the resonance is also shown as a function of proper time $\tau$.  
Right: $\omB=0.2\omega$. Then, the motion has $|\bar\vu|\ll|\Delta\vu|$ and remains regular (doubly-periodic in $\xi$, with fast $\omega$ and slow $\omL=\omega_B/\bg$). The large number of $\omega$-oscillations shown in the figure ($N=600$) merge, forming the black stripes, whose thickness demonstrates the oscillation amplitude $|\Delta\vu|\sim a_0$. The $\omega$-oscillations of  $u_\xi$ are small (the red stripe is thin).  
}
\label{fig:test}
 \end{figure*}

Fig.~\ref{fig:test} shows sample solutions demonstrating two types of motion: $|\bar\vu|\gg |\Delta \vu|$ (found when $\omega<\omB$) and $|\bar\vu|\ll |\Delta \vu|$. Both solutions were calculated for the same wave ($a_0=30$, $\xirise=150(2\pi/\omega)$), but with different $\boldsymbol{\omB}=(0,10\omega,0)$ and $\boldsymbol{\omB}=(0,\omega/5,0)$.

\medskip

{\em Waves with $\omega<\omB$.---}
In this case, the gyration of $\bar\vu$ in $\bBbg$ develops a huge amplitude $|\bar{\vu}|\approx\bg \gg |\Delta \vu|\sim a_0$, and the particle's motion becomes dominated by Larmor rotation with frequency $\omL=\omB/\bg \ll\omega$ (Fig.~\ref{fig:test}, left). We observe that $|\bar\vu|$ is pumped in nearly impulsive events that occur every Larmor rotation. These events are resonances where the particle exchanges energy with one wave oscillation $\delta\xi\sim\omega^{-1}$. 

The resonance may be described as follows. The wave oscillation along the particle's worldline resembles an oscillator with a changing frequency $\omw=(1-\beta_z)\omega$. It becomes slowest near moment $t_0$ where $\beta_z\approx 1$ is maximum and $u_\xi\approx 0$ is minimum (then the particle moves almost together with the wave). Gyration of $\vu\approx \bar\vu$ gives $\beta_z \approx \cos \delta\psi$ and $\omw\approx \omega\, (\delta \psi)^2/2$, where $\delta\psi=\omL\delta t$ and $\delta t=t-t_0$. The resonance occurs when
\beq
   \omw\delta t \sim 1 
   \quad \Rightarrow \quad 
   \delta\psi_{\rm res}= \omL\delta t_{\rm res}\sim(\omL/\omega)^{1/3}. 
\eeq
The obtained $\delta\psi_{\rm res}$  determines the characteristic $u_\xi=\gamma(1-\beta_z)$ and $u_x$ during the resonance,
\beq
\label{eq:uxi_res}
 u_\xi^{\rm res} \sim \left(\frac{\gamma\omB^2}{\omega^2}\right)^{1/3}
  \;\;\, \frac{u_x^{\rm res}}{a_0}\sim \left(\frac{\gamma}{\gs}\right)^{2/3} 
      \;\;  \gamma_\star\equiv\sqrt{\frac{a_0^3\omega}{\omB}}
\eeq
These expressions assume $u_x^{\rm res}>a_0$ which requires $\gamma>\gs$ (then $u_x^{\rm res}\approx U_x^{\rm res}\approx  -\gamma\,\sin\delta\psi_{\rm res}$). One can verify that $\omL/\omega=(\gs/\gamma)(\Bbg/E_0)^{3/2}\ll 1$ and $\delta\psi_{\rm res}\ll 1$.

\begin{figure}[t]
\vspace*{-1.5mm}
\includegraphics[width=0.45\textwidth]{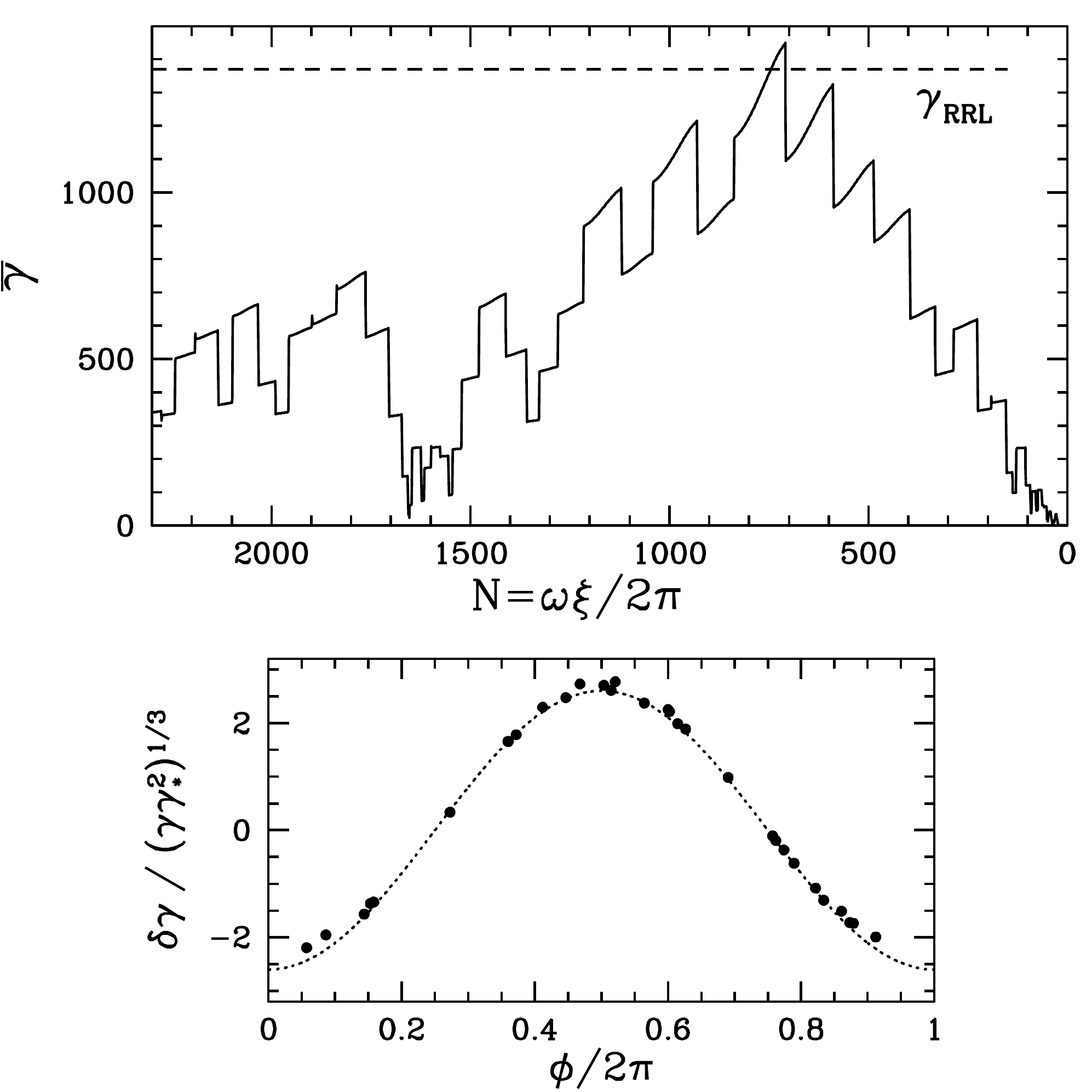} 
\caption{Same model as in Fig.~1 ($\omB=10\omega$), but now with radiation reaction. We chose $r_e\omega/c=10^{-4}\omB^2/a_0^5\omega^2$, which gives $\gL\approx 1370$ (horizontal dashed line). The evolution of $\bar{\gamma}$ consists of stochastic jumps $\delta\gamma$ (resonances) followed by gradual losses. Bottom: $\delta\gamma$ vs. wave phase $\phi$ at the resonance. The result confirms \Eq~(\ref{eq:dgam}) with $H\approx 2.6$ (dotted curve).}
\label{fig:losses}
 \end{figure}

 Gain $\delta\gamma$ from the resonance event  may be found from $d\gamma/d\tau=eEu_x/mc$ and $du_\xi/d\tau=-\omega_Bu_x$, which gives
\beq
   \frac{d\gamma}{du_\xi}=-\frac{E}{\Bbg}, \quad 
   \delta\gamma=-\frac{E_0}{\Bbg}\int \sin(\omega\delta\xi+\phi)\,du_\xi.
\eeq  
Here $\phi$ is the (practically random) phase of the wave at the particle's location at time $t_0$. Note that $u_\xi$ is even in $\delta\xi$ during the resonance, near the minimum $u_\xi$. Hence, the odd part of $\sin(\omega\delta\xi+\phi)$, i.e. $\sin(\omega\delta\xi)\cos\phi$, determines the integral and $\delta\gamma \sim  - u_\xi^{\rm res} (E_0/\Bbg)\cos\phi$. This result may be stated as
\beq
\label{eq:dgam}
\delta\gamma \approx -H\left(\gamma\gamma_\star^2\right)^{1/3}\cos\phi  \qquad  
    (\gamma>\gamma_\star).
\eeq
Exact integration gives the coefficient $H\approx 2.6$ (Fig.~\ref{fig:losses}). If the particle approaches the resonance with $\gamma<\gs$, it gains $\delta\gamma\sim\gs$. As the resonances repeat every $\xiL\approx \tL=2\pi/\omL$, the particle performs random walk in $\gamma$ to $\gamma\gg\gs$. The wave acts as a stochastic ``pump'' that can accelerate the particle to arbitrary high $\gamma$ (limited only by radiative losses discussed below). The deterministic particle motion gives the chaotic walk in $\gamma$ because $\delta\gamma$ is sensitive to $\phi$. It remains regular in repeating resonances every gyration.

The presented calculation assumes $\theta=\pi/2$ (wave propagation perpendicular to $\bBbg$). Similar integration of \Eq~(\ref{eq:dyn}) at $\theta<\pi/2$ also gives pumping of $\gamma$, with additional sliding of the particle along the oblique $\bBbg$. One can boost the reference frame along $\bBbg$ so that the wave propagates perpendicular to $\bBbg$ in the new frame $K'$, and see that the pump works as described above, with $\omega'/\omB'=(\omega/\omB)\sin\theta$. The pre-wave motion of the particle in the boosted frame is unimportant unless $\theta\ll 1$.

\medskip

{\em Waves with $\omega>\omB$.---} 
In this case, the particle moves with $|\bar\vu|\ll|\Delta\vu|$ in slowly rising waves, $\xirise\gg a_0/\omB$. The resulting motion at $\xi>\xirise$ is periodic, with no pumping effect (Fig.~\ref{fig:test}, right). 

$|\bar\vu|$ can be derived analytically. Let us average the energy equation $mc d\gamma/d\tau=eEu_x$ over the $\omega$-oscillations:
\beq
\label{eq:bg1}
    u_\xi \frac{d\bg}{d\xi}\approx \frac{e}{mc}\left(\overline{EU_x}-\overline{Ea}\right)\approx \frac{d}{d\xi}\frac{\baa}{2}.
\eeq
Here we used $eEa/mc=-a\,da/d\xi$ and neglected  $\overline{EU_x}=\overline{U_xE_0\sin(\omega\xi)}$ since $U_xE_0$ varies slowly and $\overline{\sin(\omega\xi)}=0$. Dynamical equations for $u_x$ and $u_z$ may be written as one equation for complex $u=u_x+iu_z$. After averaging, it becomes
\beq
\label{eq:bu}
   \frac{d\bu}{d\tau}\approx i\omega_B\bu+iF, \qquad F\equiv\frac{1}{2}\frac{d\baa}{d\xi}\approx \frac{1}{4}\frac{da_0^2}{d\xi}.
\eeq
The Larmor rotation of $\bu$ is excited where the wave rises, $F\neq 0$, and the solution of \Eq~(\ref{eq:bu}) during the rise is  
\beq
\label{eq:bu1}
   \bu\approx i \int_0^{\tau} e^{i\omega_B(\tau-\tau')} F(\tau')\,d\tau'\approx -\frac{F}{\omB} \quad (\xi<\xirise),
\eeq 
where second equality uses the slow-rise approximation (we integrated by parts and neglected $dF/d\tau\ll \omB F$). Then, from $du_\xi/d\tau\approx -\omega_B\bux\approx F\approx d\bg/d\tau$, we find $u_\xi\approx \bg$, and \Eq~(\ref{eq:bg1}) yields 
$\bg^2\approx 1+a_0^2/2$. At $\xi>\xirise$, 
\beq
\label{eq:small_omB}
   \bg \approx \sqrt{1+\frac{a_0^2}{2}},   \quad
   \bu\approx \frac{a_0^2\,e^{i\psi}}{2\omB\xirise},
   \quad u_\xi\approx \bg-|\bu|\sin\psi,
\eeq
where $\psi=\omB(\tau-\tau_{\rm rise})$. This analytical result agrees with the numerical solution.

One can also evaluate the integral in \Eq~(\ref{eq:bu1}) when $\xirise\ll a_0/\omB$. Then, we find periodic motion at $\xi>\xirise$ with $|\bar\vu|\sim (a_0^3/\omega_B \xirise)^{1/2}\gg |\Delta\vu|\sim a_0$. 

\medskip

{\em Radiation reaction limit (RRL).---}
A relativistic electron in fields $\bE$ and $\bB+\bBbg$ emits momentum with rate \citep{Landau75}
\beq
\label{eq:RR}
   \dgem mc
  = \frac{\sT}{4\pi}\left\{ [\gamma\bE+\vu\times(\bB+\bBbg)]^2-(\vu\cdot\bE)^2\right\}.
\eeq
We now retain $\bfrr=-\dgem m\bv$ in the dynamical~\Eq~(\ref{eq:dyn}). For waves with $E_0\gg\Bbg$, $\dgem mc$ simplifies to $\sT E^2u_\xi^2/4\pi$, and  averaging over $\omega$-oscillations gives
\beq
\label{eq:dgem}
  \overline{\dot{\gamma}}_{\rm em}\approx \frac{r_e}{3c}\,a_0^2\omega^2 u_\xi^2, \qquad r_e=\frac{e^2}{mc^2}.
\eeq
  
\begin{figure}[t]
\vspace*{-2.26cm}
\hspace*{0.3cm}
\includegraphics[width=0.43\textwidth]{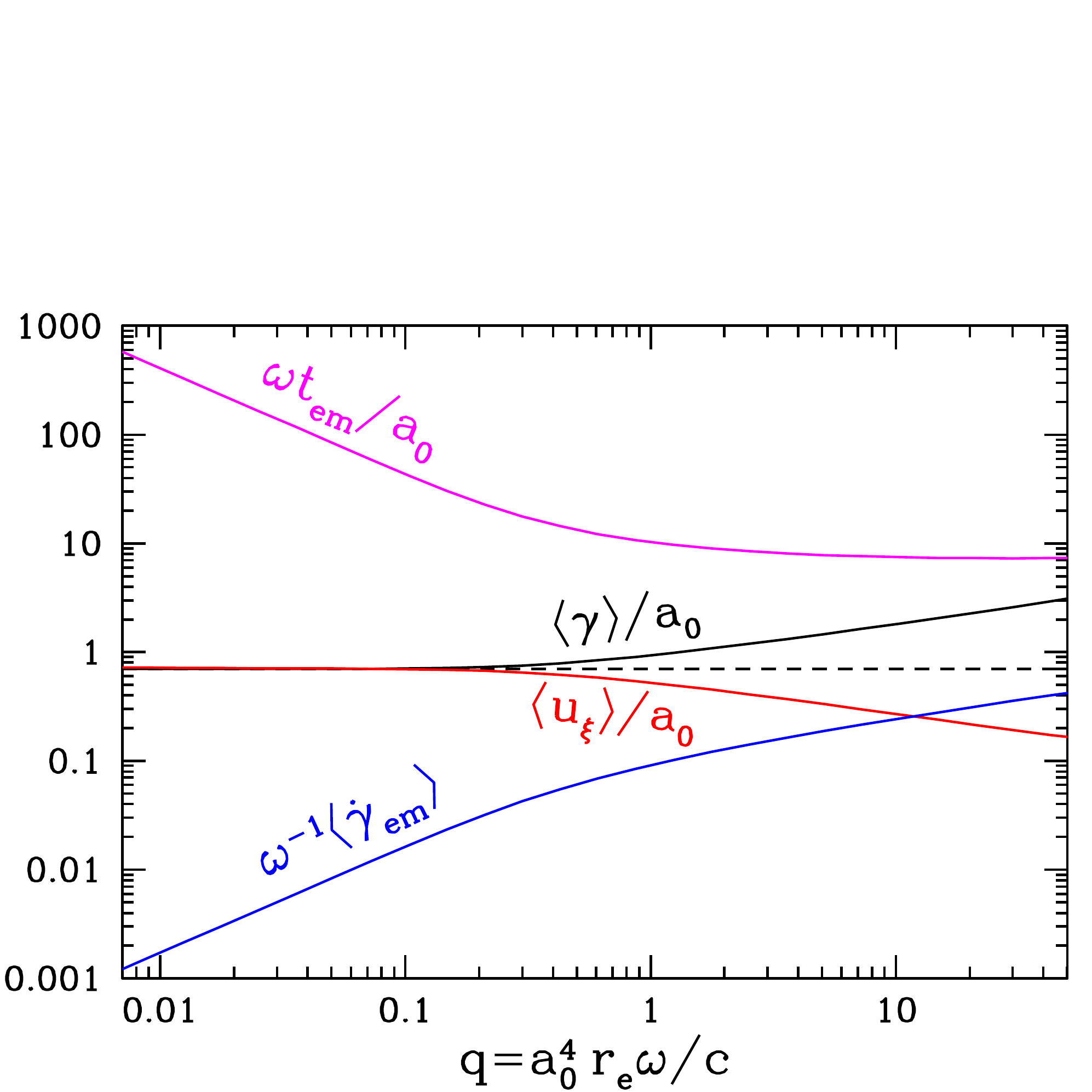} 
\caption{
Development of radiation reaction with increasing $q$ in waves with $\omega>\omB$. Averaging $\langle...\rangle$ is performed over a time longer than the particle gyration in $\Bbg$; $t_{\rm em}\equiv \gav/\langle\dot{\gamma}_{\rm em}\rangle$. The plot was constructed by solving a sequence of models with varying $r_e\omega/c$ at fixed $a_0=30$ and $\omB=0.1\omega$, however the same result holds for other choices of $a_0\gg 1$ and $\omB<\omega$.
}
\label{fig:losses_q}
 \end{figure}

In waves with $\omega<\omB$, the resonant pumping of $\gamma$ quickly reaches RRL (Fig.~\ref{fig:losses}), and random walk continues with a ceiling $\gL$. Losses occur with $\langle u_\xi^2\rangle =2\langle \gamma^2\rangle$ (averaged over gyration), and balance the maximum resonant gain $\delta\gamma=H(\gamma\gs^2)^{1/3}$ when $\langle\dgem\rangle \tL = \delta\gamma$. This gives (using \Eq~\ref{eq:dgem})
\beq
\label{eq:gL}
   \gL =\gs \left(\frac{a_\star}{a_0}\right)^{15/8} 
     =\left(\frac{3H}{4\pi}\frac{c}{r_e\omega a_0}\right)^{3/8}
     \left(\frac{\omB}{\omega}\right)^{1/4}
\eeq
This result holds if $\gL>\gs$, i.e. if
\beq
\label{eq:astar}
  a_0<a_\star \equiv  \left(\frac{3Hc\, \omB^2}{4\pi r_e\omega^3}\right)^{{1}/{5}}
  \approx 400\,\nu_{\rm GHz}^{-1/5} \left(\frac{\omB}{\omega}\right)^{2/5}
\eeq
where $\nu=\omega/2\pi$ is normalized to 1\,GHz. (If $a_0>a_\star$, losses completely suppress diffusion in $\gamma$.) 

Timescale for reaching RRL is $t_{\rm RRL}\sim (\gL/\delta\gamma)^2\tL$, which gives $\omega t_{\rm RLL}/2\pi \sim H^{-2}(E_0/\Bbg)^{3/2}(\as/a_0)^{35/8}$. For a bright FRB, $t_{\rm RRL}$ is shorter than the FRB duration ($\sim 1\,$ms), so the wave pushes particles to the RRL.

When $\omega>\omB$, we find that radiation reaction is negligible if $q\equiv a_0^4r_e\omega/c\ll 1$; then the particle keeps $\gamma\sim a_0$. If $q>1$ then $\gamma$ {\em grows} (Fig.~\ref{fig:losses_q}), because $u_\xi$ develops $\omega$-oscillations and radiative losses become asymmetric in phase, inducing a rocket effect.

\begin{figure}[t]
\vspace*{-3.45cm}
\hspace*{0.2cm}
\includegraphics[width=0.45\textwidth]{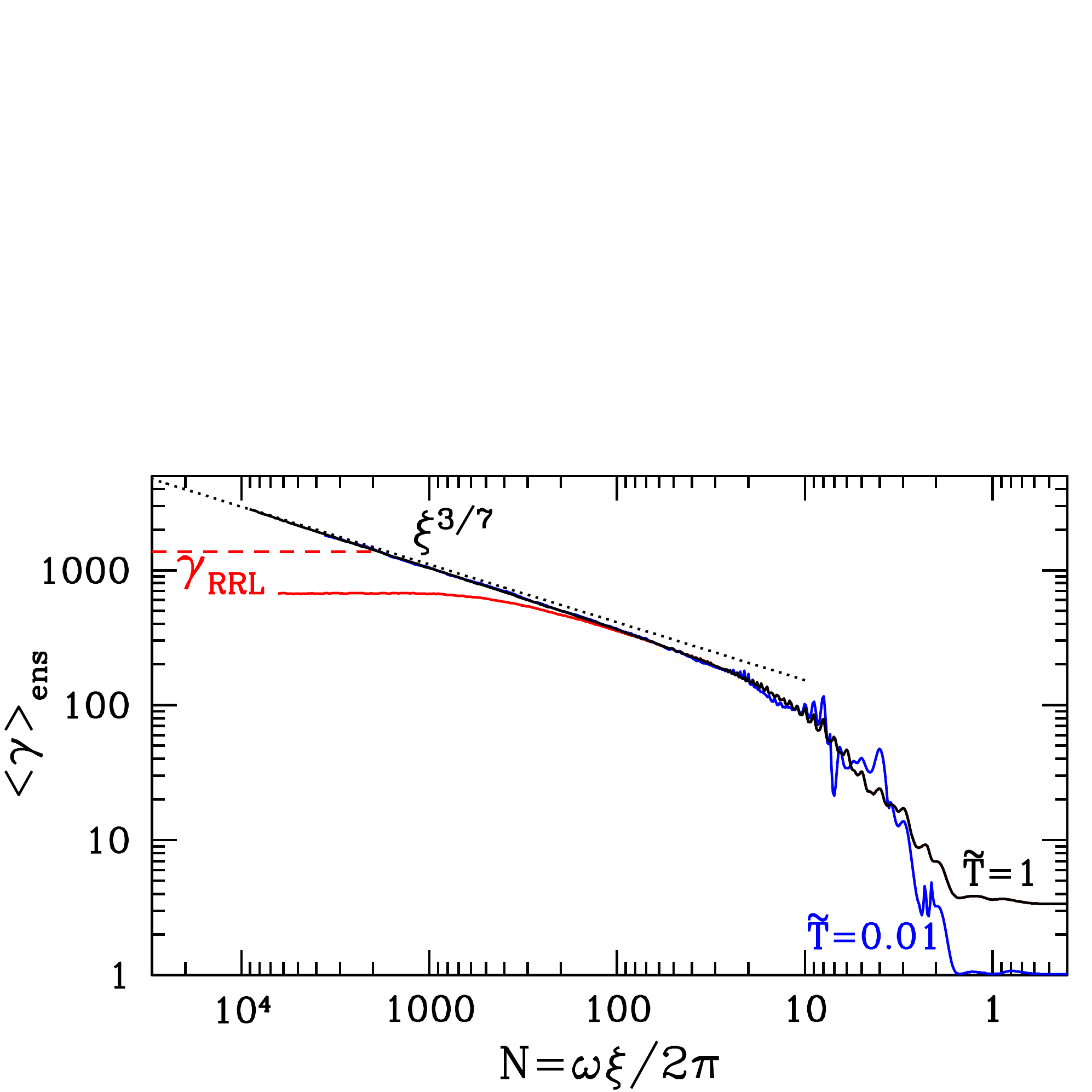} 
\caption{Wave pumps $\langle\gamma\rangle_{\rm ens}$ of a particle ensemble with initial $T\neq 0$ ($a_0=30$, $\omega\xirise/2\pi=10$, $\omB=10\omega$). Three models are shown: with no radiative losses for $\tilde{T}\equiv kT/mc^2=0.01$ (blue) and 1 (black), and with losses for $\tilde{T}=1$, $\gL=1370$ (red). Black dotted line shows the acceleration law $\langle\gamma\rangle_{\rm ens}\propto \xi^{3/7}$.}
\label{fig:gamma_av}
 \end{figure}

\medskip

{\em Initial temperature $T\neq 0$.---} 
A simple way to see the statistics of chaos realizations in a wave packet with $\omega<\omB$ is to draw an ensemble of test particles from an initial Maxwellian distribution with some $T\neq 0$. We followed $10^4$ particles and observed evolution of their distribution function $f(\gamma)$ in the wave. Fig.~\ref{fig:gamma_av} shows the evolution of ensemble average $\langle\gamma\rangle_{\rm ens}$ for three models with $kT/mc^2=0.01$ and 1. As  chaos develops, $\langle\gamma\rangle_{\rm ens}$ is pumped to high values, losing memory of initial $T$. The observed growth of $\langle\gamma\rangle_{\rm ens}\propto \xi^{3/7}\propto t^{3/7}$ is consistent with the simple diffusion estimate of the characteristic $\gamma^2\sim D t$ where $D(\gamma)\approx \langle [\delta\gamma]^2\rangle/\tL\propto \gamma^{-1/3}$ (\Eq~\ref{eq:dgam}). Radiative losses offset acceleration at $\langle\gamma\rangle_{\rm ens}\approx\gL/2$.

\medskip

{\em Scattering cross section $\ssc$.---}
Time-averaged emitted power $\dEem=\langle \dgem\rangle mc^2$ determines the scattering cross section of the particle $\ssc=\dEem/F$, where $F=c E_0^2/8\pi$ is the wave energy flux. \Eq~(\ref{eq:dgem}) gives $\ssc\approx \langle u_\xi^2\rangle\,  \sT$. If $\Bbg=0$, the particle keeps $u_\xi=1$, so $\ssc=\sT$ (and in frame $K'$, where $\bu_z'=0$, $\ssc'/\sT=1+3a_0^2/8$ \citep{Landau75}).

It is particularly interesting to look at $\ssc$ for $\omega<\omB$. Then, $\langle u_\xi^2\rangle\sim \gL^2$. A characteristic $\ssc^\star$ may be defined with $a_0=\as$:
\beq
   \frac{\ssc^\star}{\sT} \sim
   \left(\frac{c}{r_e\omega}\right)^{3/5}\left(\frac{\omega_B}{\omega}\right)^{1/5} 
   \sim 10^8\,\nu_{\rm GHz}^{-3/5} \left(\frac{\omega_B}{\omega}\right)^{1/5}.
\eeq
Recall that this result holds for $E_0>\Bbg$. If $E_0$ is reduced below $\sim \Bbg/2$,
$\ssc$ would drop to $(\omega^2/\omB^2)\sT$.

\medskip

{\em Energies of emitted photons.---}
The emitted power $\dgem mc^2$ is carried by curvature radiation with spectrum extending to a characteristic frequency $\omsc\approx (3/2) \gamma^3 c/r_c$, where $r_c^{-1}=(3\dEem/2cr_e\gamma^4)^{1/2}$ is the curvature of the particle trajectory \citep{Landau75}. Substitution of \Eq~(\ref{eq:dgem}) gives
\beq
   \frac{\omega_c}{\omega}\approx a_0\gamma u_\xi.
\eeq
When $\omega>\omB$, we find $\langle\gamma u_\xi\rangle\sim a_0^2$ (Fig.~\ref{fig:losses_q}) and $\omega_c\sim a_0^3\,\omega$.

For $\omega<\omB$, we use $\langle \gamma u_\xi\rangle \approx 2\gL^2$ and \Eq~(\ref{eq:gL}) to get
\beq
    \frac{\hbar\omsc}{mc^2}
        \approx  \frac{1}{\alpha} \left(\frac{r_e\omB^2 a_0}{c\,\omega}\right)^{1/4}
        = \epsc^\star \left(\frac{a_0}{\as}\right)^{1/4},
\eeq
\beq
 \epsc^\star\approx   \frac{1}{\alpha}
   \left(\frac{r_e\omB^3}{c\,\omega^2}\right)^{1/5}
    \approx 0.3 \left(\frac{\omB}{\omega}\right)^{3/5} \nu_{\rm GHz}^{1/5},
 \eeq
where $\alpha=e^2/\hbar c$. Waves with $\omega\ll\omB$ generate photons with $\hbar\omega_c>m_ec^2$ capable of $e^\pm$ creation.

\medskip

{\em Discussion.---}
Strong low-frequency waves (\Eq~\ref{eq:cond}) offer a novel mechanism for particle acceleration near astrophysical compact objects. It differs from stochastic acceleration by plasma turbulence, where particles gain energy from interactions with many plasma modes \citep{Petrosian12}.  The wave induces a peculiar resonance with the particle motion (without fine-tuning $\omega$) which repeats nearly impulsively at a specific gyration phase $\psi$ with a random wave phase $\phi$, giving random energy boosts to the particle. This behavior is an incidence of chaos development in nonlinear dynamics. Different examples of chaos in plasma waves with a magnetostatic background are found in \citep{Smith75,Karney77,Menyuk87,Chen01}; chaotic motion in electrostatic waves was particularly well studied  \citep{Sagdeev88}.  Ultrastrong radio waves in regime~(\ref{eq:cond}) provide a remarkable new example, which admits simple description presented in this Letter. 

Reproducing this acceleration mechanism in the lab is difficult. \citep{Apollonov88} considered particle acceleration in a laser beam with $a_0\gtrsim 1$ propagating across a static $\bBbg$ with a Larmor radius $r_{\rm L}=c/\omL$ exceeding the beam size. Reaching a small $r_{\rm L}$ and the conditions~(\ref{eq:cond}) or (\ref{eq:astar}) is difficult because of limited $\Bbg$ accessible to experiments. Another  experimental setup engineers a slow wave (phase speed $\vph<c$) trapping particles at wave phases $\phi$ where $E>|\bB-\bBbg|$ \citep{Takeuchi87,Yugami96}. This surfatron accelerator is not realized in a neutron star magnetosphere (the radio waves have $\vph\geq c$). Instead, stochastic acceleration described in this Letter results from many short resonances with random $\phi$, repeated every Larmor rotation in $\Bbg$.

Strong waves accelerate protons as well as electrons. The RRL energy scales with the particle mass as $m^{3/2}$, however reaching this limit takes time $t_{\rm RRL}\propto m^{7/2}$. Therefore, ion acceleration (to be studied in future work) will be limited by exposure to the wave rather than $t_{\rm RRL}$. Future work should also extend our calculations to non-planar waves with a finite beaming angle $\theta_b$; we expect the plane-wave approximation to hold if $\theta_b<\psi_{\rm res}$.

The quick acceleration of electrons in a strong radio wave has important astrophysical implications, which will be investigated in detail elsewhere. Curvature emission with $\hbar\omsc>m_ec^2$ will lead to an $e^\pm$ avalanche capable of   powering observed X-ray bursts from magnetars. Magnetar quakes first excite low-frequency Alfv\'en waves, whose nonlinear interactions generate strong radio waves in the magnetosphere \citep{Thompson98,Troischt04,Li19}. Our results suggest that these waves do not silently escape, as usually assumed. Instead, they will generate powerful $e^\pm$ fireworks in the outer magnetosphere where $\Bbg\lesssim E_0$. Similar waves are expected in a magnetized neutron star binary before its merger, and the resulting $e^\pm$ fireworks may be observed as an X-ray precursor of the merger.
 
Strong implications are inevitable for FRB models that picture a bright GHz source near a magnetar. The accompanying paper \citep{Beloborodov21}  shows that the FRB will experience enormous scattering in the outer magnetosphere, failing to pass through radii $R=10^9$-$10^{10}\,$cm. This implies that observable FRBs must be emitted by relativistic ejecta from the magnetosphere.

The analysis of particle dynamics in ultrastrong waves in this Letter assumed that the  wave propagates with the vacuum speed $\vph=c$ (\Eq~\ref{eq:EB}), neglecting any collective plasma effects on the propagation speed. Collective effects (in particular wave dispersion, $\vph \neq c$) are discussed in \citep{Beloborodov21}. In main applications, dispersion turns out negligible compared to wave damping due to scattering by individual particles, which is an interesting special feature of ultrastrong waves. In particular, FRBs are choked by scattering in a plasma of modest density, when deviations of $\vph$ from $c$ are still negligible.

\medskip

This work is supported by NSF grant AST 2009453, Simons Foundation grant \#446228, and the Humboldt Foundation.

\bibliography{scat}

\end{document}